\documentclass[aps,prl,twocolumn,preprintnumbers,superscriptaddress,10pt,footinbib]{revtex4-1}

\usepackage[mathscr]{eucal}

\usepackage{epsfig,latexsym,amsfonts,amsmath,amsthm,amssymb,amsbsy,multirow,slashed,wasysym,textcomp,subfigure,hyperref,wrapfig,datetime,comment,mathtools,cancel}

 \newcommand{\badat}{\begin{alignedat}}
 \newcommand{\eadat}{\end{alignedat}}

\newcommand{\eal}[1]{\be \begin{aligned} #1 \end{aligned}\end{equation}} 
\newcommand{\eqn}[1]{\be #1 \end{equation}} 
\newcommand{\eqa}[1]{\bea  #1\end{eqnarray}}

\long\def\new#1\endnew{{\bf #1}}		
\long\def\del#1\enddel{}

\def\del{\partial}

\def\A{\mathcal{A}}

\def\K{\mathcal{K}}

\def\tPhi{\widetilde \Phi}
\def\tn{\widetilde n}

\def\bz{{\bar z}}

\def\eps{\epsilon}

\def\ma{\mathfrak{a}}
\def\mb{\mathfrak{b}}

\begin{document}

\preprint{CPHT-RR049.072020}

\title{Double Copy for Celestial Amplitudes}

\author{Eduardo Casali}
\affiliation{\it Center for Quantum Mathematics and Physics (QMAP) and Department of Physics, University of California, Davis, CA 95616 USA}

\author{Andrea Puhm}
\affiliation{\it CPHT, CNRS, Ecole polytechnique, IP Paris, F-91128 Palaiseau, France}

\begin{abstract}
  \noindent
Celestial amplitudes which use conformal primary wavefunctions rather than plane waves as external states offer a novel opportunity to study properties of amplitudes with manifest conformal covariance and give insight into a potential holographic celestial CFT at the null boundary of asymptotically flat space. Since translation invariance is obscured in the conformal basis, features of amplitudes that heavily rely on it appear to be lost. Among these are the remarkable relations between gauge theory and gravity amplitudes known as the double copy. Nevertheless, properties of amplitudes reflecting fundamental aspects of the perturbative regime of quantum field theory are expected to survive a change of basis. Here we show that there exists a well-defined procedure for a  \textit{celestial double copy}. This requires a generalization of the usual squaring of numerators which entails first promoting them to generalized differential operators acting on external wavefunctions, and then squaring them. We demonstrate this procedure for three and four point celestial amplitudes, and give an argument for its validity to all multiplicities.
\end{abstract}

\maketitle

\section{Introduction}

Scattering amplitudes are usually calculated using asymptotic states in a plane waves basis. 
Applying a Mellin transform to the energy of each external state in an amplitude amounts to a change in basis of the asymptotic states from plane waves to socalled {\it conformal primary wavefunctions}~\cite{Pasterski:2017kqt}. In this basis, the four-dimensional Lorentz group $SL(2,\mathbb{C})$ acts manifestly as the group of conformal transformations of the {\it celestial} two-sphere at null infinity. Amplitudes in which asymptotic states are in this conformal basis make conformal covariance manifest~\cite{Pasterski:2016qvg,Pasterski:2017ylz,Schreiber:2017jsr,Stieberger:2018edy,Strominger:2013lka,Strominger:2013jfa,Strominger:2017zoo} and thus offer a novel opportunity to study properties of amplitudes that may be obscured in a plane wave basis.  
Moreover, it is hoped that studying these {\it celestial amplitudes} will shed light on aspects of a possible flat space holographic principle, where a CFT living on the celestial sphere would be dual to a bulk theory living on asymptotically flat spacetime.

Celestial amplitudes have a rather complicated analytic structure already at tree-level which is further complicated by the fact that the Mellin transform mixes the UV and the IR. It is therefore curious yet reassuring that universal aspects of amplitudes such as soft factorization theorems survive in the conformal basis~\cite{Cheung:2016iub,Donnay:2018neh,Fan:2019emx,Pate:2019mfs,Adamo:2019ipt,Puhm:2019zbl,Guevara:2019ypd,Law:2019glh,Fotopoulos:2019vac,Banerjee:2020kaa,Fan:2020xjj}.
In general, features of amplitudes that are expected to reflect fundamental properties of the perturbative regime of quantum field theory should survive a change of basis. 

One such remarkable feature are the double copy relations~\cite{Bern:2010ue}, which state that gravitational amplitudes can be obtained by a well-defined ``squaring'' of gauge theory amplitudes. These relations are known to hold to all multiplicities at tree-level~\cite{BjerrumBohr:2009rd,Stieberger:2009hq,Feng:2010my,Bern:2010yg} and can be seen as implied by string theory relations~\cite{Plahte:1970wy,Kawai:1985xq,BjerrumBohr:2010hn,BjerrumBohr:2010zs,BjerrumBohr:2009rd,Stieberger:2009hq,Tourkine:2016bak,Hohenegger:2017kqy,Casali:2019ihm,Casali:2020knc,Vanhove:2018elu,Vanhove:2020qtt} in their low energy limit. They also hold at loop-level in a plethora of cases and for many pairs of field theories (see~\cite{Bern:2019prr} for a comprehensive review). Moreover, there are versions of the double copy relating full non-linear solutions of gauge theory and gravity~\cite{Monteiro:2014cda,Monteiro:2015bna,Luna:2018dpt,Luna:2015paa,Luna:2016hge,Berman:2018hwd,Ridgway:2015fdl,DeSmet:2017rve,Bahjat-Abbas:2017htu,Carrillo-Gonzalez:2017iyj,Lee:2018gxc,Gurses:2018ckx,Huang:2019cja,Alawadhi:2019urr,Banerjee:2019saj}.
Thus it is reasonable to expect the double copy to also hold in some form after a change of basis for the asymptotic particles.

However, because translation invariance is obscured in the conformal basis, a generalization of the double copy that does not rely on momentum conservation is required.
This is not dissimilar to the situation for amplitudes in curved backgrounds for which a notion of the double copy was shown to persist~\cite{Adamo:2017nia,Adamo:2017sze,Adamo:2018mpq,Adamo:2020syc,Adamo:2020qru,Adamo:2019zmk}.
This provides another incentive to study the double copy for celestial amplitudes as it presents aspects of curved space amplitudes without the complications of non-trivial backgrounds.

Here we propose a well-defined procedure for a \textit{celestial double copy}. This requires a generalization of the usual squaring of numerators to first promoting them to differential operators acting on external scalar wavefunctions and then squaring them. 
Moreover, our proposal offers an interesting presentation for gauge and gravity amplitudes as operators acting on scalar amplitudes.

\section{Background}\label{sec:Review}

We consider massless scattering processes in four-dimensional Minkowski space $\mathbb{R}^{1,3}$, with spacetime coordinates $X^\mu$ for $\mu=0,1,2,3$, where states are labeled by null momentum four-vectors $k_\mu$, such that $k^2=0$.
\\ \vspace{-0.3cm}

\textbf{Momentum Basis vs Conformal Basis.} 
Scattering problems of gauge bosons are usually studied in the plane wave basis which consists of spin-one wavefunctions
$ \epsilon_{\mu;\ell}(k)e^{i k \cdot X}$ 
in Lorenz gauge, and spin-two wavefunctions
$ \epsilon_{\mu\nu;\ell}(k)e^{i k \cdot X}$ 
in harmonic (de Donder) gauge.
Here $\epsilon_{\mu;\ell}(k)$ are the polarization vectors for helicity $\ell$ one-particle states which satisfy $\epsilon_\ell(k)\cdot k=0$, $\epsilon_{\pm}(k)^*=\epsilon_{\mp}(k)$, and $\epsilon_\ell(k)\cdot \epsilon_{\ell'}(k)^*=\delta_{\ell \ell'}$. The polarization tensor is taken to be $\epsilon_{\mu\nu;\ell}(k)=\epsilon_{\mu;\ell}(k)\epsilon_{\nu;\ell}(k)$. Plane wave solutions are thus labeled by the spatial momentum $\vec{k}$, the four-dimensional helicity $\ell$ and a sign distinguishing incoming from outgoing states. 

To make conformal symmetry manifest an alternative basis of massless wavefunctions was constructed in~\cite{Pasterski:2017kqt}, called \textit{conformal primary wavefunctions}. These are labeled by a point $(z,\bz)$ on the celestial sphere and the quantum numbers under the conformal group, namely the conformal dimension $\Delta$ and the two-dimensional spin $J$, as well as a sign $s$ distinguishing incoming $(-)$ from outgoing $(+)$ states. 
Writing the momentum four-vector
\begin{equation}
 k_\mu=s \omega q_\mu(z,\bz)\,,
\end{equation}
massless spin-zero conformal primaries satisfying the scalar wave equation are related to the plane wavefunctions by a Mellin transform~\footnote{An $i\varepsilon$ prescription needs to be introduced to circumvent the singularity at the light sheet where $q\cdot X=0$. }
\begin{equation}\label{phiDelta}
 \phi^{\Delta,\pm}(X^\mu;z,\bz)=\int_0^\infty d\omega \omega^{\Delta-1} e^{\pm i \omega q\cdot X}=\frac{(\mp i)^\Delta \Gamma(\Delta)}{(-q\cdot X)^\Delta}\,.
\end{equation}
We construct massless spin-one wavefunctions as
\begin{equation}\label{Vspin1}
 V^{\Delta,\pm}_{\mu;J}(X^\mu;z,\bz)=\epsilon_{\mu;J}(z,\bz) \phi^{\Delta,\pm}(X^\mu;z,\bz)\,,
\end{equation}
and spin-two wavefunctions as
\begin{equation}\label{Vspin2}
 V^{\Delta,\pm}_{\mu\nu;J}(X^\mu;z,\bz)= \epsilon_{\mu \nu;J}(z,\bz) \phi^{\Delta,\pm}(X^\mu;z,\bz)\,,
\end{equation}
where we identified the two-dimensional spin $J$ with the four-dimensional helicity $\ell$, and the polarization vectors of, respectively, helicity $+1$ and $-1$ one-particle states propagating in the $q^\mu$ direction, are $\partial_z q^\mu=\sqrt{2}\epsilon^\mu_+(q)$ and $\partial_\bz q^\mu=\sqrt{2}\epsilon^\mu_-(q)$. The polarization tensors of helicity $+2$ and $-2$ states are obtained by the product of same helicity polarization vectors.
The wavefunctions~\eqref{Vspin1}-\eqref{Vspin2} satisfy, respectively, the Maxwell and linearized Einstein equations and, up to pure gauge terms, transform covariantly under $SL(2,C)$ Lorenz transformations and as two-dimensional conformal primaries with, respectively, spin $J=\pm1$ and $J=\pm2$ and conformal dimension $\Delta$~\cite{Cheung:2016iub,Pasterski:2017kqt}.
In~\cite{Pasterski:2017kqt} they were shown to form a complete $\delta$-function normalizable basis for $\Delta\in 1+i\mathbb{R}$ on the principal continuous series of the $SL(2,\mathbb C)$ Lorentz group~\footnote{A subtlety arises for zero-modes with $\Delta=1$ which was addressed in~\cite{Donnay:2018neh}.}.
\\ \vspace{-0.3cm}

\textbf{Celestial Amplitudes.} 
Given an amplitude in the plane wave basis 
\begin{equation}
 \A_{n}(\{\omega_j,\ell_j;z_j,\bz_j\})=A_{n} \, {\delta^{(4)}\Big(\sum_{j=1}^n k^\mu_j\Big)}\,,
\end{equation}
the celestial amplitude is obtained by a Mellin transform on each of the external particles as
\begin{equation} \label{CelesitalAmplitude}
 \widetilde{\A}_{n}(\{\Delta_j,J_j;z_j,\bz_j\})=\prod_{j=1}^n \Big(\int_0^\infty d\omega_j \omega_j^{\Delta_j-1}\Big) \A_{n}\,.
\end{equation}
Here $k^\mu_j=s_j\omega_j q^\mu_j(z_j,\bz_j)$ are the momenta of the massless particles with $\omega_j> 0$ and null vectors $q^\mu_j$ directed at points $(z_j,\bz_j)$ on the celestial sphere at null infinity where the particles cross, with $s_j=\pm$ labeling outgoing/incoming states.
In~\cite{Pasterski:2017ylz,Pasterski:2017kqt,Stieberger:2018edy} celestial amplitudes were shown to share conformal properties with correlation functions on the celestial sphere. 
At the same time translation invariance becomes obscured in the conformal basis. Hence, properties of amplitudes that make heavy use of momentum conservation appear to be lost in celestial amplitudes. An example of this is the perturbative double copy relating Einstein gravity amplitudes to the ``square'' of Yang-Mills amplitudes~\cite{Bern:2019prr}.
\\ \vspace{-0.3cm}

\textbf{Double Copy.} 
Let $\Gamma_n$ be the set of trivalent graphs with $n$ external legs and no internal closed loops, corresponding to tree-level amplitudes~\footnote{This representation also holds for loop integrands, but we restrict ourselves here to tree-level amplitudes.}. Yang-Mills amplitudes can be presented as a sum over these trivalent graphs as
\begin{equation}\label{eq:YM_tree}
 \mathcal{A}^{YM}_n=\delta^{(4)}\Big(\sum_{j=1}^n k^\mu_j\Big)\sum_{\gamma\in\Gamma_n} \frac{c_\gamma n_\gamma}{\Pi_\gamma}\,,
\end{equation}
where, for a particular trivalent graph $\gamma$, the numerators $c_\gamma$ are its color factors, that is, traces of the Lie algebra matrices of the external particles. The numerators $n_\gamma$ are the kinematical factors given by polynomials in the momenta $k_j$ and polarizations $\epsilon_j$ of the external particles. The denominator $\Pi_\gamma$ is a product of the scalar propagators associated with the graph $\gamma$. The color factors $c_\gamma$ are not all independent due to the Lie algebra Jacobi identity. For a triple of graphs related by a BCJ move~\cite{Bern:2008qj}, the color factors obey $ c_s-c_t+c_u=0$.
At tree-level we can always find kinematical numerators $n_\gamma$ that obey the same identities as the color factors
$ n_s-n_t+n_u=0$.
These are called color-kinematics dual numerators, or BCJ satisfying numerators~\cite{Bern:2008qj}. Substituting the color factors in \eqref{eq:YM_tree} by these numerators we obtain
\begin{equation}\label{eq:G_tree}
 \A^{G}_n=\delta^{(4)}\Big(\sum_{j=1}^n k^\mu_j\Big)\sum_{\gamma\in\Gamma_n} \frac{n_\gamma^2}{\Pi_\gamma}\,,
\end{equation}
which turns out to be an amplitude for external gravitons - this is the double copy procedure~\cite{Bern:2010yg}. 
This presentation of the double copy relies heavily on the fact that external particles are in the plane wave basis and on explicit momentum conservation. Both of these require a generalization in order to arrive at a double copy for celestial amplitudes.

\section{Celestial Double Copy}\label{sec:CDC}

Consider a representation of the amplitudes obtained from position space Feynman diagrams where spacetime integrals like
\begin{equation}\label{deltaInt}
 \delta^{(4)}\Big(\sum_{j=1}^nk_j^\mu\Big)=\int \frac{d^4 X}{(2\pi)^4} e^{\sum_{j=1}^n i k_j \cdot X}
 \end{equation} 
are left undone.
Rewriting the Yang-Mills amplitude~\eqref{eq:YM_tree} and the gravitational amplitude~\eqref{eq:G_tree} using the integral represenation~\eqref{deltaInt} gives rise to the most natural representation in which to generalize the double copy, as first discussed in~\cite{Adamo:2017nia}.

To compute celestial gluon and graviton amplitudes we use as external states the conformal wavefunctions~\eqref{Vspin1}-\eqref{Vspin2}, namely 
\begin{equation}\label{eq:V_states}
 V_\mu(X)=T^\ma\epsilon_\mu\phi(X)\,, \quad 
 V_{\mu\nu}(X)=\epsilon_{\mu}\epsilon_{\nu}\phi(X)  \,,
\end{equation}
where we dropped various labels and added a color index.
In analogy to how spacetime derivatives act on wavefunctions in a momentum basis we define the generalized celestial momentum $K_\mu$ via $ \partial_\mu\phi(X)= K_\mu(X)\phi(X)$ as
\begin{equation}\label{eq:CelestialMomentum}
 K_\mu(X)=\frac{\Delta q_\mu}{(-q\cdot X)}\,.
\end{equation}
The explicit dependence of $K_\mu$ on spacetime is to be contrasted with the spacetime independent $k_\mu$ of the plane wave basis and is a precursor for the failure of the naive squaring procedure for celestial amplitudes. 
\\ \vspace{-0.3cm}

\textbf{Three-Point Amplitudes.} 
From the Yang-Mills Lagrangian we extract the three-point vertex
\begin{equation}\label{eq:YM_vertex}
\badat{2}
 \widetilde{\mathcal{A}}^{YM}_3=&\frac{1}{2}\text{Tr}\int \frac{d^4X}{(2\pi)^4}\,(V_1^\mu V_2^\nu\partial_\nu V_{3\mu}-V_2^\mu V_1^\nu\partial_\nu V_{3\mu}\\
&\qquad\qquad\qquad\qquad\qquad\qquad\quad+\text{cyclic})\,,
 \eadat
\end{equation}
and for the gravitational amplitude we use
\begin{equation}\label{eq:G_vertex}
\badat{2}
\widetilde{\mathcal{A}}^{G}_3&=\frac{1}{2}\int \frac{d^4X}{(2\pi)^4}\,(V_1^{\mu\nu}\partial_\mu V_{2\rho\sigma}\partial_\nu V_3^{\rho\sigma}\\
&\qquad-2 V_1^{\rho\nu}\partial_\mu V_{2\rho\sigma}\partial_\nu V_3^{\mu\sigma}+\text{ permutations})
\eadat
\end{equation}
as three-point interaction evalutated in the conformal basis. This is most easily obtained from the action introduced in~\cite{Cheung:2016say,Cheung:2017kzx} which was checked in the plane wave basis in~\cite{Adamo:2017nia} to match the three-point interaction from the Einstein-Hilbert Lagrangian.
Inserting the external states~\eqref{eq:V_states} into~\eqref{eq:YM_vertex} and~\eqref{eq:G_vertex} we obtain the celestial three-gluon amplitude
\begin{equation}\label{CelestialYMV3}
\badat{1}
 \widetilde{\A}^{\rm YM}_{3}
 &=f^{\ma_1\ma_2\ma_3}\frac{1}{2}\int\frac{d^4X}{(2\pi)^4}\left((\epsilon_1\cdot (K_2-K_3)\; \epsilon_2\cdot \epsilon_3 \right. \\ &\quad\quad\quad\quad\quad\quad\quad\quad\quad\quad\quad\quad \left. +\text{cyclic}\right)\prod_{j=1}^3\phi_j\,,
\eadat
\end{equation}
and the celestial three-graviton amplitude
\begin{equation}\label{CelestialgravityV3}
\badat{2}
 &\widetilde{\mathcal{A}}^{G}_{3}
 =\frac{1}{2}\int\frac{d^4X}{(2\pi)^4} \;\left((\epsilon_2\cdot \epsilon_3)^2 \;\epsilon_1 \cdot K_2\;\epsilon_1 \cdot K_3\right.\\ &\left. -2\,\epsilon_1 \cdot \epsilon_2\;\epsilon_2 \cdot \epsilon_3 \;\epsilon_3\cdot K_2\;\epsilon_1 \cdot K_3  +\text{perm.}\right)\prod_{j=1}^3\phi_j\,.
\eadat
\end{equation}
In the plane wave basis we would proceed by squaring the Yang-Mills integrand and use momentum conservation $\sum_j k^\mu_j=0$, as well as $\epsilon_j \cdot k_j=0$ to show equality with the gravity integrand. In the celestial case, the conformal wavefunctions still obey $\epsilon_j \cdot K_j=0$ due to our choice of gauge, but because the celestial momenta are spacetime dependent momentum conservation is no longer manifest in~\eqref{CelestialYMV3}-\eqref{CelestialgravityV3}~\footnote{While in~\eqref{CelestialYMV3} we still have $\sum_{j=1}^3 K_j^\mu(X)\simeq 0$ via integration by parts, this is no longer true after squaring. }.
A naive double copy of the celestial Yang-Mills numerator $(\tn_{YM})^2=\frac{1}{4}\left((\epsilon_1\cdot (K_2-K_3)\; \epsilon_2\cdot \epsilon_3 +\text{cyclic}\right)^2$
gives terms proportional to $(\epsilon_i \cdot \epsilon_j)^2$ 
as well as cross-terms
which one might hope to reexpress as the corresponding terms in the gravity integrand~\eqref{CelestialgravityV3} via integration by parts.
However, due to the spacetime dependence of the celestial momenta, performing integration by parts on terms quadratic in the same momenta produces extra terms which are not present in the gravity amplitude resulting in an obstruction to the naive double copy. 
\\ \vspace{-0.3cm}

\noindent\underline{\textit{Celestial double copy:}} Our proposal is to introduce formal differential operators $\K_i$ that replace the generalized celestial momenta $K_i$. These operators are defined to act on the conformal wavefunctions $\phi_i$ as
\begin{equation}\label{TranslationOperator}
 \K_i^\mu \phi_j(X) = 
 K_i^\mu(X) \phi_i(X) \delta_{ij}\,.
\end{equation}
Notice that these operators keep track of particle labels and thus are \textit{not} equivalent to the usual partial derivatives. Instead they admit a representation as~\footnote{We thank Atul Sharma for pointing out this relation.}
\begin{equation}\label{eq:diff_external}
 \K^\mu_i = q^\mu_i e^{\partial_{\Delta_i}} \,,
\end{equation}
which is the translation generator of the Poincar\'{e} algebra acting on the $i$-th particle~\cite{Stieberger:2018onx}.
Promoting the celestial momenta $K_i$ in~\eqref{CelestialYMV3} to the operators $\K_i$ and squaring gives
\begin{equation}\label{MellinYMsquare}
\badat{3}
 (\mathcal{N}_{YM})^2&=\frac{1}{4}(\epsilon_1\cdot (\K_2-\K_3) (\epsilon_2\cdot \epsilon_3)+\text{cyclic})^2\,,
\eadat
\end{equation}
which acts on the external wavefunctions in the three-point amplitude. 
Evaluating this action explicitly yields terms that differ from the naive double copy by factors of $1+\frac{1}{\Delta_i}$ in terms with the same celestial momenta. These arise from
\begin{equation}
 \K_i^\mu \K_i^\nu \phi_i(X) = \left(1+\frac{1}{\Delta_i}\right) K_i^\mu(X) K_i^\nu(X) \phi_i(X) \,,
\end{equation}
and cancel precisely the additional terms generated when integrating by parts~\eqref{MellinYMsquare}, matching it to the gravitational integrand.
The final result is
\begin{equation}\label{CelestialYMV3square}
\badat{2}
&(\mathcal{N}_{YM})^2\prod_{i=1}^3\phi_i =-(\epsilon_2\cdot \epsilon_3)^2 \;\epsilon_1 \cdot K_2\;\epsilon_1 \cdot K_3\\ &+2 \;\epsilon_1 \cdot \epsilon_2\;\epsilon_2\cdot \epsilon_3 \;\eps_3 \cdot K_2\;\eps_1 \cdot K_3 +\text{cyclic})\prod_{i=1}^3\phi_i \,,
\eadat
\end{equation}
which is, up to a sign, exactly the gravitational numerator in~\eqref{CelestialgravityV3}. Hence, the celestial three-gluon amplitude double copies into the celestial three-graviton amplitude.
\\ \vspace{-0.3cm}

\textbf{Four-Point Amplitudes.}
Here, the new ingredients are the presence of the four-point contact term and the exchange diagrams 
\begin{equation}\label{eq:4_amplitude}
 \widetilde{\mathcal{A}}^{YM}_4= \widetilde{\mathcal{A}}^{s}_{4}+\widetilde{\mathcal{A}}^{t}_{4}+\widetilde{\mathcal{A}}^{u}_{4}+\widetilde{\mathcal{A}}^{contact}_{4}\,.
\end{equation}
As in the plane wave basis, we split the contact term into three contributions according to the color factors, using that the propagator 
\begin{equation}\label{eq:propagator}
 G_{\mu\nu}(X,Y)=\eta_{\mu\nu}\,G(X,Y)=\eta_{\mu\nu}\int d^4 k \frac{e^{i k\cdot(X-Y)}}{k^2}
\end{equation}
satisfies $\Box_X G(X,Y)=(2\pi)^4 \delta^{(4)}(X-Y)=\Box_Y G(X,Y)$ as well as the fact that the celestial momenta are null, $K_i^2(X)=0$.
This yields
\begin{multline}\label{eq:BCJ_amp}
 \widetilde{\mathcal{A}}^{YM}_4=\int \frac{d^4X}{(2\pi)^4} \frac{d^4Y}{(2\pi)^4}G(X,Y)\\
 \times\left(c_s \tn_s\tPhi_s+c_t \tn_t\tPhi_t+c_u \tn_u\tPhi_u\right)\,,
\end{multline}
where
\begin{equation}\label{eq:ctPhi}
 \begin{array}{ccc}\centering
  c_s=f^{\ma_1\ma_2\mb}f^{\ma_3\ma_4\mb}\,,\quad & \tPhi_s=\phi_1\phi_2(X)\phi_3\phi_4(Y)\,,\\
  c_t=f^{\ma_1\ma_3\mb}f^{\ma_2\ma_4\mb}\,,\quad & \tPhi_t=\phi_1\phi_3(X)\phi_2\phi_4(Y)\,,\\
  c_u=f^{\ma_1\ma_4\mb}f^{\ma_2\ma_3\mb}\,,\quad & \tPhi_u=\phi_1\phi_4(X)\phi_2\phi_3(Y) \,,
 \end{array}
\end{equation}
and the numerators are given by
\begin{multline}\label{eq:s_numerator_celestial}
 \tn_s=[\epsilon_1\cdot\epsilon_2\;(K_1-K_2)^\mu + 2\;\epsilon_1\cdot K_2\;\epsilon_2^\mu - 2\;\epsilon_2\cdot K_1\;\epsilon_1^{\mu}](X)\;\eta_{\mu\nu}\\
 \times[\epsilon_3\cdot\epsilon_4\;(K_4-K_3)^{\nu}-2\;\epsilon_3\cdot K_4\;\epsilon_4^\nu + 2\;\epsilon_4\cdot K_3 \;\epsilon_3^\nu](Y)\\ + (\epsilon_1\cdot\epsilon_3\;\epsilon_2\cdot\epsilon_4-\epsilon_1\cdot\epsilon_4\;\epsilon_2\cdot\epsilon_3)(K_1\cdot K_2(X)+K_3\cdot K_4(Y))\,,
\end{multline}
with appropriate permutations for the other channels.

Following the three-point case, we promote all celestial momenta $K_i$ to operators acting on the scalar wavefunctions
\begin{multline}\label{eq:s_op_numerator}
 \mathcal{N}_s=[\epsilon_1\cdot\epsilon_2\;(\K_1-\K_2)^\mu + 2\epsilon_1\cdot \K_2\;\epsilon_2^\mu - 2\;\epsilon_2\cdot \K_1\;\epsilon_1^{\mu}]\;\eta_{\mu\nu}\\
 \times[\epsilon_3\cdot\epsilon_4\;(\K_4-\K_3)^{\nu}-2\;\epsilon_3\cdot \K_4\;\epsilon_4^\nu + 2\;\epsilon_4\cdot \K_3 \;\epsilon_3^\nu]\\ + (\epsilon_1\cdot\epsilon_3\;\epsilon_2\cdot\epsilon_4-\epsilon_1\cdot\epsilon_4\;\epsilon_2\cdot\epsilon_3)(\K_1\cdot \K_2+\K_3\cdot \K_4)\,.
\end{multline}
The gravity amplitude is obtained by substituting each color factor by its corresponding kinematical operator
\begin{multline}\label{eq:BCJ_amp_c}
 \widetilde{\mathcal{A}}^{G}_4=\int \frac{d^4X}{(2\pi)^4} \frac{d^4Y}{(2\pi)^4}G(X,Y)\\
 \times\left((\mathcal{N}_s)^2\tPhi_s+(\mathcal{N}_t)^2\tPhi_t+(\mathcal{N}_u)^2\tPhi_u\right)\,.
\end{multline}
To check it, we rely on the fact that any set of numerators of a trivalent graph decomposition of a four-point Yang-Mills amplitude in the plane wave basis, 
\begin{multline}\label{eq:BCJ_amp_pw}
 \A^{YM}_4=\int \frac{d^4X}{(2\pi)^4} \frac{d^4Y}{(2\pi)^4}\,G(X,Y)\\\left(c_sn_s\Phi_s+c_tn_t\Phi_t+c_un_u\Phi_u\right)\,,
\end{multline}
obeys color-kinematics and, hence, the Yang-Mills numerators square to a gravitational amplitude
\begin{multline}\label{eq:BCJ_usual}
 \A^{G}_4=\int \frac{d^4X}{(2\pi)^4} \frac{d^4Y}{(2\pi)^4}\,G(X,Y)\\\left((n_s)^2\Phi_s+(n_t)^2\Phi_t+(n_u)^2\Phi_u\right)\,.
\end{multline}
Here $n_s$ and $\Phi_s$ are given by expressions analogous to, respectively,~\eqref{eq:ctPhi} and~\eqref{eq:s_numerator_celestial} with $K_j$ replaced by $k_j$ and $\phi_j$ replaced by $e^{ik_j\cdot X}$.
To show that~\eqref{eq:BCJ_amp_c} is indeed the Mellin transform of the gravitational amplitude~\eqref{eq:BCJ_usual} note that it depends  polynomially on the energy of each particle $k^0_i$. There are two cases to consider: the numerator is linear in the energy
\begin{equation}
 \int d \omega_i \omega_i^{\Delta_i-1} (ik_i^\mu) e^{ik_i\cdot X}=K_i^\mu\phi_i(X)\,,
\end{equation}
and the numerator is quadratic in the same energy
\begin{equation}
 \int d \omega_i \omega_i^{\Delta_i-1} (ik_i^\mu) (ik_i^\nu) e^{ik_i\cdot X}=\left(1+\frac{1}{\Delta_i}\right)K_i^\mu K_i^\nu\phi_i(X)\,.
\end{equation}
Since this mirrors the action of the $\K_i$ operators on the conformal wavefunctions, Mellin transforming~\eqref{eq:BCJ_usual} gives precisely~\eqref{eq:BCJ_amp_c} after the action of the $\K_i$ operators. Hence the celestial four-gluon amplitude double copies into the celestial four-graviton amplitude.

\section{Discussion}\label{sec:Discussion}

The n-point generalization is straightforward. Given a set of color-kinematics satisfying numerators promote them to operator valued numerators as delineated above. These are then squared as \textit{operators} and act on the respective scalar amplitude. This gives the representations
\begin{equation}\label{eq:BCJ_cel}
\tilde A^{YM}_n=\sum_{\gamma\in\Gamma}c_\gamma\mathcal{N}_{\gamma}S_\gamma\rightarrow  \tilde A^G_n=\sum_{\gamma\in\Gamma}(\mathcal{N}_{\gamma})^2S_\gamma \,,
\end{equation}
where $\Gamma$ is the set of trivalent graphs and $S_\gamma$ is the scalar amplitude for the trivalent graph $\gamma$. It is clear that this is the correct gravity amplitude since the construction above corresponds to the Mellin transform of a double copied amplitude by the same arguments used at four points. Thus our prescription is valid to arbitrary multiplicity. 

To illustrate this we write down explicitly the four-point amplitudes in this representation:
\begin{equation}
\badat{2}
 \widetilde{\mathcal{A}}^{YM}_4&=c_s\;\mathcal{N}_s\;\mathfrak{s}+c_t\;\mathcal{N}_t\;\mathfrak{t}+c_u\;\mathcal{N}_u\;\mathfrak{u}\,,\\
 \widetilde{\mathcal{A}}^{G}_4&=\left(\mathcal{N}_s\right)^2\mathfrak{s}+\left(\mathcal{N}_t\right)^2\mathfrak{t}+\left(\mathcal{N}_u\right)^2\mathfrak{u}\,,
 \eadat
\end{equation}
where the s-channel contribution to the cubic four-point scalar amplitude is
\begin{equation}\label{eq:scalar_s}
 \mathfrak{s}=\int \frac{d^4X}{(2\pi)^4} \frac{d^4Y}{(2\pi)^4}G(X,Y)\prod_{i=1}^2 \phi_i(X) \prod_{j=3}^4 \phi_j(Y)\,,
\end{equation}
with $\mathfrak{t}$ and $\mathfrak{u}$ obtained by the corresponding permutations. 
It would be interesting to understand how to obtain these numerators directly in the celestial framework instead of building on their plane wave representations, and what is the celestial analogue of the Jacobi relations obeyed by color-kinematics satysifying numerators. This could also hint at a geometric interpretation of this kinematical algebra as found in~\cite{Monteiro:2011pc} for the MHV sector.

The way we presented celestial amplitudes is reminiscent of recent results on Ambitwistor strings in Anti-de Sitter spacetime~\cite{Roehrig:2020kck,Eberhardt:2020ewh} which also has operator-valued kinematical numerators. The difference here is that propagators are generated by the usual Feynman rules while the Ambitwistor string uses the scattering equations to do so. 
Since Ambitwistor string theories also provide formulas with manifest double copy structure~\cite{Mason:2013sva} it would be natural to study them in the celestial context, and in fact, they have already been useful in studying some properties of celestial amplitudes~\cite{Adamo:2019ipt,Adamo:2014yya,Adamo:2015fwa}.

We expect that the natural way to implement the double copy in {\it curved} spacetimes will be through operator-valued numerators. Our work thus presents an important stepping stone in this direction. It would also be interesting to see how our proposal for the celestial double copy translates into a statement about the two-dimensional correlators directly in the putative celestial CFT.

\section*{Acknowledgements}
We would like to thank Tim Adamo, Agnese Bissi, Cindy Keeler, Sebastian Mizera, Sabrina Pasterski and Piotr Tourkine for discussions, and the Centro de Ciencias de Benasque Pedro Pascual where this work was initiated for its hospitality. This research is supported in part by U.S. Department of Energy grant DE-SC0009999, funds provided by the University of California and the European Research Council (ERC) under the European Union’s Horizon 2020 research and innovation programme (grant agreement No 852386).

\bibliography{references}

\end{document}